\documentclass[aps,twocolumn,superscriptaddress,prb,floatfix,british]{revtex4}

\usepackage{amssymb}
\usepackage{amsmath}
\usepackage{babel}
\usepackage{graphicx}
\usepackage{psfrag}
\usepackage[sort&compress]{natbib}
\usepackage{topcapt}
\usepackage{hyperref}

\newcommand{\ket}[1]{\left\lvert #1 \right\rangle}
\DeclareMathOperator{\tr}{Tr}

\newcommand{\Sapphire}{\mathrm{Al}_2 \mathrm{O}_3}
\newcommand{\Hethree}{^3\mathrm{He}}
\newcommand{\Hefour}{^4\mathrm{He}}

\newcommand{\mK}{\mathrm{mK}}

\newcommand{\MHz}{\mathrm{MHz}}
\newcommand{\GHz}{\mathrm{GHz}}
\newcommand{\us}{\mu\mathrm{s}}
\newcommand{\ns}{\mathrm{ns}}

\newcommand{\nm}{\mathrm{nm}}
\newcommand{\um}{\mu \mathrm{m}}
\newcommand{\mm}{\mathrm{mm}}

\newcommand{\EJmax}{E_{\mathrm{J}}^{\mathrm{max}}}

\newcommand{\EJqmax}{E_{\mathrm{J}q}^{\mathrm{max}}}
\newcommand{\EC}{E_{\mathrm{C}}}

\newcommand{\ECq}{E_{\mathrm{C}q}}
\newcommand{\EJq}{E_{\mathrm{J}q}}

\newcommand{\rhoML}{\rho_{\mathrm{ml}}}

\newcommand{\QL}{Q_{\mathrm{L}}}
\newcommand{\QR}{Q_{\mathrm{R}}}
\newcommand{\FreqL}{f_{\mathrm{L}}}
\newcommand{\FreqR}{f_{\mathrm{R}}}
\newcommand{\VL}{V_{\mathrm{L}}}
\newcommand{\VR}{V_{\mathrm{R}}}

\newcommand{\VH}{V_{\mathrm{H}}}
\newcommand{\FC}{f_{\mathrm{C}}}

\newcommand{\wcav}{\omega_{\mathrm{C}}}
\newcommand{\wcavontpi}{\omega_{\mathrm{C}}/2\pi}

\newcommand{\Rxp}{R_x^{\pi\protect\vphantom{/2}}}

\newcommand{\Rypt}{R_y^{\pi/2}}

\newcommand{\Rtomo}{R_{x,y}^{0,\pi/2,\pi}}

\newcommand{\cPhij}{\mathrm{c}U_{ij}}
\newcommand{\cPhz}{\mathrm{c}U_{00}}
\newcommand{\cPho}{\mathrm{c}U_{01}}
\newcommand{\cPhtw}{\mathrm{c}U_{10}}
\newcommand{\cPhth}{\mathrm{c}U_{11}}

\begin{document}
\title{Demonstration of Two-Qubit Algorithms with a Superconducting Quantum Processor}
\author{L.\ DiCarlo}
\affiliation{Departments of Physics and Applied Physics, Yale University, New Haven, CT 06511, USA}
\author{J.\ M.\ Chow}
\affiliation{Departments of Physics and Applied Physics, Yale University, New Haven, CT 06511, USA}
\author{J.\ M.\ Gambetta}
\affiliation{Department of Physics and Astronomy and Institute for Quantum Computing, University of Waterloo, Waterloo, Ontario N2L 3G1, Canada}
\author{Lev\ S.\ Bishop}
\affiliation{Departments of Physics and Applied Physics, Yale University, New Haven, CT 06511, USA}
\author{B.\ R.\ Johnson}
\affiliation{Departments of Physics and Applied Physics, Yale University, New Haven, CT 06511, USA}
\author{D.\ I.\ Schuster}
\affiliation{Departments of Physics and Applied Physics, Yale University, New Haven, CT 06511, USA}
\author{J.\ Majer}
\affiliation{Atominstitut der \"{O}sterreichischen Universit\"{a}ten, TU-Wien, A-1020 Vienna, Austria}
\author{A.\ Blais}
\affiliation{D\'{e}partement de Physique, Universit\'{e} de Sherbrooke, Sherbrooke, Qu\'{e}bec J1K 2R1, Canada}
\author{L.\ Frunzio}
\affiliation{Departments of Physics and Applied Physics, Yale University, New Haven, CT 06511, USA}
\author{S.\ M.\ Girvin }
\affiliation{Departments of Physics and Applied Physics, Yale University, New Haven, CT 06511, USA}
\author{R.\ J.\ Schoelkopf}
\affiliation{Departments of Physics and Applied Physics, Yale University, New Haven, CT 06511, USA}
\date{May 1, 2009}
\maketitle

\textbf{
By harnessing the superposition and entanglement of physical states, quantum computers could outperform
their classical counterparts in solving problems of technological impact, such as factoring large numbers and searching databases~\cite{Nielsen00,Kaye07}. A quantum processor executes algorithms by applying a programmable sequence of gates to an initialized register of qubits, which coherently evolves into a final state containing the result of the computation. Simultaneously meeting the conflicting requirements of long coherence, state preparation, universal gate operations, and qubit readout makes building quantum processors challenging.
Few-qubit processors have already been shown in nuclear magnetic resonance~\cite{Chuang98a,Jones98,Chuang98b,Vandersypen01}, cold ion trap~\cite{Guide03,Brickman05} and optical~\cite{Kwiat00} systems, but a solid-state realization has remained an outstanding challenge.
Here we demonstrate a two-qubit superconducting processor and the implementation of the Grover search~\cite{Grover97} and Deutsch--Jozsa~\cite{Deutsch92} quantum algorithms. We employ a novel two-qubit interaction, tunable in strength by two orders of magnitude on nanosecond time scales, which is mediated by a cavity bus in a circuit quantum electrodynamics (cQED) architecture~\cite{Blais04,Wallraff04}.
This interaction allows generation of highly-entangled states with concurrence up to $\mathbf{94}\boldsymbol{\%}$.
Although this processor constitutes an important step in quantum computing with integrated circuits, continuing efforts to increase qubit coherence times, gate performance and register size will be required to fulfill the promise of a scalable technology.
}

Over the last decade, superconducting circuits~\cite{Clarke08} have made considerable progress on all the
requirements necessary for an electrically-controlled, solid-state quantum computer.
Coherence times~\cite{Clarke08,Schreier08}  have risen by three orders of magnitude to $\sim 1\,\us$, single-qubit gates~\cite{Lucero08,Chow09} have reached error rates of
$1\%$, engineered interactions~\cite{Yamamoto03,Hime06,Plantenberg07,Niskanen07} have produced two-qubit entanglement at a level of $60\%$ concurrence~\cite{Steffen06}, and qubit readout~\cite{Siddiqi04,McDermott05,Steffen06} has attained measurement fidelities $\sim 90\%$. However, combining these achievements in a single device
remains challenging. One approach to integration is the quantum bus architecture~\cite{Blais04,Sillanpaa07,Majer07}, which uses an on-chip transmission line cavity to couple, control, and measure qubits.  We augment the architecture in Ref.~\onlinecite{Majer07} with flux-bias lines that tune individual qubit frequencies, permitting single-qubit phase gates.
By pulsing the qubit frequencies to an avoided crossing where a $\sigma_{z}\otimes\sigma_{z}$ interaction turns on, we are able to realize a two-qubit conditional phase (c-Phase) gate. Operation in the strong-dispersive regime~\cite{Schuster07} of cQED allows joint readout~\cite{Filipp08} that can efficiently detect two-qubit correlations. Combined with single-qubit rotations, this enables tomography of the two-qubit state. Through an improved understanding of spontaneous emission~\cite{Houck08} and careful microwave engineering, we are now able to combine state-of-the-art $\sim1\,\us$ coherence times into a two-qubit device. This allows sufficient time to concatenate $\sim10$ gates, realizing simple algorithms with fidelity greater than
$80\%$.

\begin{figure*}[htbp!]
%\psfrag{c}[c][c][1.0]{$\FC$}
%\psfrag{d}[c][c][1.0]{$\FreqR$}
%\psfrag{e}[c][c][1.0]{$\FreqL$}
%\psfrag{f}[c][c][1.0]{$\VL$}
%\psfrag{g}[c][c][1.0]{$\VR$}
%\psfrag{h}[c][c][1.0]{$\FC$}
%\psfrag{n}[c][c][1.0]{$\VH$}
\centering
\includegraphics[width=6.5in]{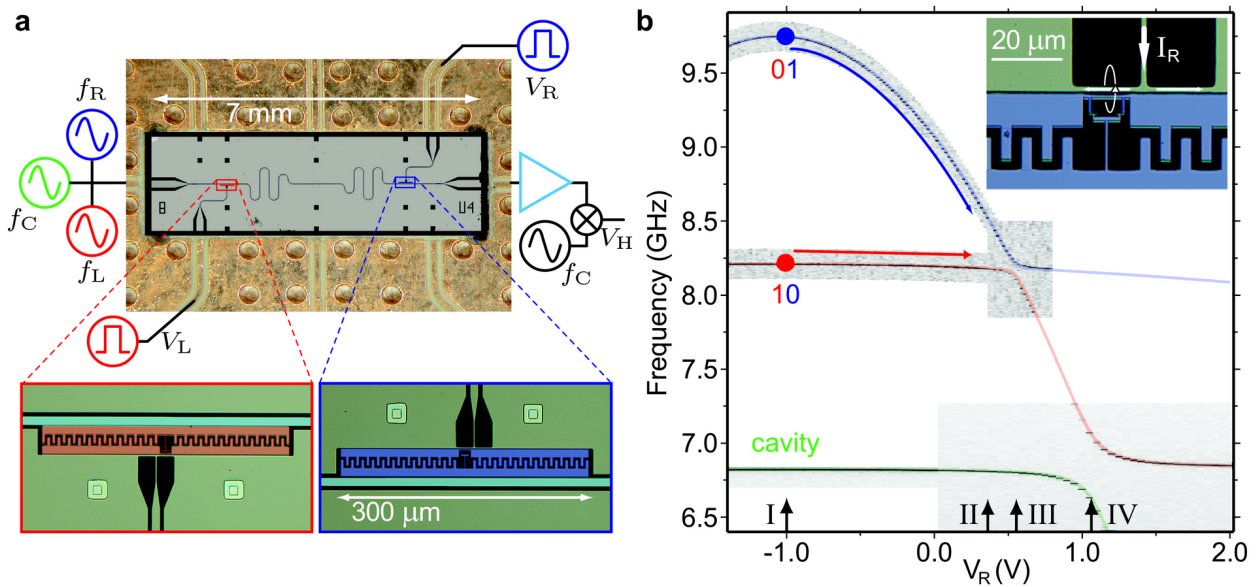}
\caption{{\bfseries Two-qubit cQED device, and cavity/qubit characterization}.
{\bfseries a},~Optical micrograph of 4-port device with a coplanar waveguide cavity  bus coupling two transmon qubits (insets), and local flux-bias lines providing fast qubit tuning. Microwave pulses at the qubit transition frequencies $\FreqL$ and $\FreqR$ drive single-qubit rotations, and
a pulsed measurement of the cavity homodyne voltage $\VH$ (at frequency $\FC$) provides two-qubit readout.
The flux-bias lines (bottom-left and top-right ports)  are coplanar waveguides with short-circuit termination next to their target qubit. The termination geometry allows current on the line
to couple flux through the split junctions (b, inset). {\bfseries b},~Grey scale images of cavity transmission and of qubit spectroscopy as a function of $\VR$, showing local tuning of $\QR$ across the avoided crossing  with $\QL$ (point~III) and across the vacuum Rabi splitting with the cavity (point~IV).
Semi-transparent lines are theoretical best fits obtained from numerical diagonalization of a generalized Tavis--Cummings Hamiltonian~\cite{Tavis68}.
Points~I and~II are the operating points of the processor. Preparation, single-qubit operations and measurements are performed at point~I, and a c-Phase gate is achieved by pulsing into point~II.
\label{fig:fig1}}
\end{figure*}

Our processor, shown in Fig.~\ref{fig:fig1}a, is a 4-port superconducting device comprising two transmon qubits~\cite{Koch07,Schreier08} ($\QL$ and $\QR$) inside a microwave cavity bus, and flux-bias lines proximal to each qubit. The cavity, normally off-resonance with the qubit transition frequencies $\FreqL$ and $\FreqR$, couples the qubits by virtual photon exchange and shields them from the electromagnetic continuum. As previously demonstrated~\cite{Majer07},  microwave pulses resonant with $\FreqL$ or $\FreqR$ applied to the cavity input port provide frequency-multiplexed single-qubit $x$- and $y$-rotations with high fidelity~\cite{Chow09} and selectivity. Pulsed measurement of the homodyne voltage $\VH$ on the output port of the cavity provides qubit readout. The remaining two ports create local magnetic fields that tune the qubit transition frequencies. Each qubit has a split Josephson junction, so its frequency depends on the flux $\Phi$ through the loop according to $ h f \approx \sqrt{8\EC^{\vphantom{m}}\EJmax \lvert \cos(\Phi/\Phi_0)|}-\EC$, where $\EC$ is the charging energy, $\EJmax$ is the maximum Josephson energy, $h$ is Planck's constant, and $\Phi_0$ is the flux quantum.
By employing short-circuited transmission lines with a bandwidth from dc to $2\,\GHz$, we
can tune $\FreqL$ and $\FreqR$ by many $\GHz$ using room temperature voltages $\VL$ and $\VR$.
Static tuning of qubit transitions using the flux-bias lines is demonstrated in Fig.~\ref{fig:fig1}b.

The spectrum of single excitations (Fig.~\ref{fig:fig1}b) shows the essential features of the cavity-coupled two-qubit Hamiltonian and allows a determination of the relevant system parameters (see Methods). When the qubits are tuned to their maximum frequencies, point~I, they are far detuned from the cavity and from each other, so that interactions are small.
This point is therefore used for state preparation, single-qubit rotations and measurement, in the computational
basis $\ket{0,0}$, $\ket{0,1}$, $\ket{1,0}$, and $\ket{1,1}$,  where $\ket{l,r}$ denotes excitation level $l$ ($r$) for $\QL$ ($\QR$).
Operation at this point is also desirable because it is a flux sweet spot~\cite{Schreier08} for both qubits, providing
long coherence, with relaxation and dephasing times $T_{1,\mathrm{L(R)}}=1.3 (0.8)\,\us$ and
$T_{2,\mathrm{L(R)}}^{\ast}=1.8 (1.2)\,\us$, respectively. Tuning $\QR$ into resonance with the cavity, point~IV,
reveals a vacuum Rabi splitting~\cite{Wallraff04} from which the qubit-cavity interaction strength is extracted.
Tuning $\QR$ into resonance with $\QL$, point~III, shows an avoided crossing resulting from a cavity-mediated, qubit-qubit transverse interaction~\cite{Blais04,Blais07} investigated previously~\cite{Majer07}. In this work, we perform two-qubit gates at point~II, where no interactions are immediately apparent on examining the one-excitation manifold.\looseness=-1

\begin{figure}[htbp!]
\centering
\includegraphics[width=3.25 in]{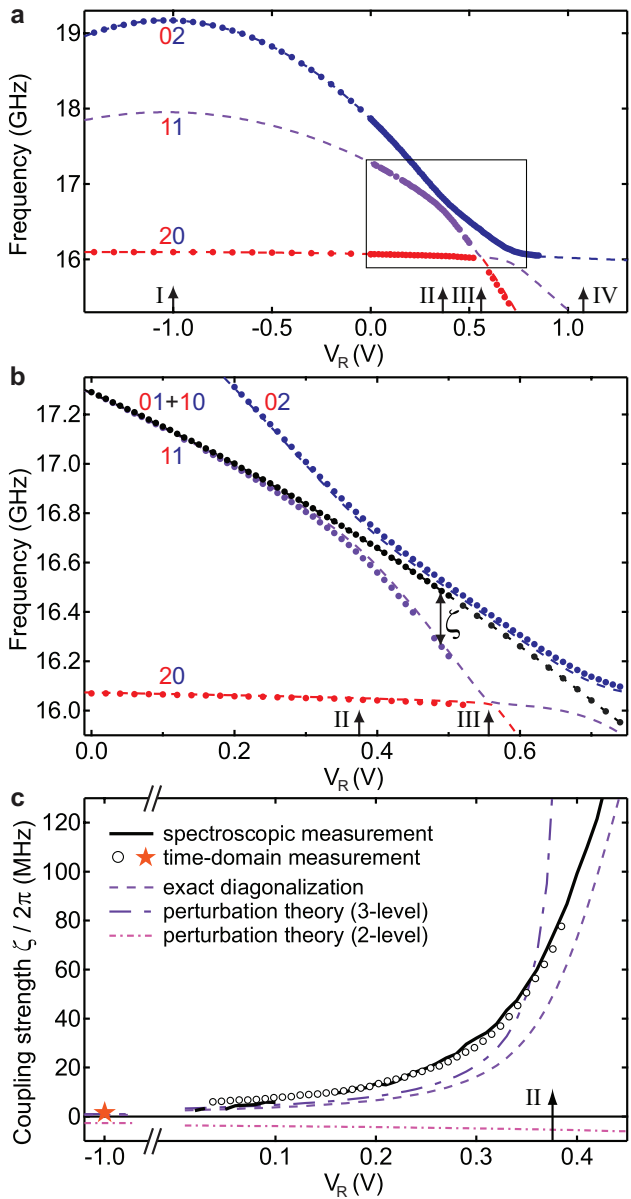}
\caption{\textbf{Origin and characterization of the controlled-phase gate.}
\textbf{a,}~Flux dependence of transition
frequencies from the ground state $\ket{0,0}$ to the two-excitation
manifold. Spectroscopy measurements (points) show an avoided crossing between the  computational
state  $\ket{1,1}$ and the non-computational state $\ket{0,2}$ at point~II, in good agreement with numerical diagonalization of the Hamiltonian (dashed curves).
\textbf{b,} This avoided crossing causes the transition frequency to $\ket{1,1}$ to deviate from the sum of the transition frequencies
to $\ket{0,1}$ and $\ket{1,0}$. \textbf{c,}~The coupling strength $\zeta/2\pi = f_{01}+f_{10}-f_{11} $ of the effective $\sigma_{z}^{\mathrm{L}} \otimes \sigma_{z}^{\mathrm{R}}$ interaction, obtained both from spectroscopy (solid curve) and from time-domain experiments (points) (see text for details).
Numerical diagonalization and perturbation theory (Supplementary Information) for 3-level transmons
agree reasonably with data. The perturbation calculation diverges at the avoided crossing.
Perturbation theory for 2-level qubits gives the wrong magnitude and sign for $\zeta$,
and demonstrates that the higher transmon excitations are necessary for the interaction.
Time-domain measurement and theory both give $\zeta/2\pi\simeq 1.2\,\MHz$ at point~I\@.
The tunability of $\zeta$ over two orders of magnitude  provides an excellent on-off ratio for the c-Phase gate.
\label{fig:fig2}}
\end{figure}

However, a useful two-qubit interaction is revealed in the two-excitation spectrum, shown in Fig.~\ref{fig:fig2}a.
As $\VR$ is swept away from point~I, the non-computational higher-level transmon excitation $\ket{0,2}$
decreases more rapidly than the computational state $\ket{1,1}$, and these states would become degenerate at point~II\@.
But as shown in Fig.~\ref{fig:fig2}b, there is a large $(160\,\MHz)$ cavity-mediated interaction between these levels, producing a frequency shift $\zeta/2\pi$ of the lower branch with respect to the sum $\FreqL+\FreqR$, in good agreement with a numerical diagonalization of the generalized Tavis--Cummings Hamiltonian~\cite{Tavis68} (see Methods).

This shift is the mechanism at the heart of our
conditional phase gate. Flux pulses, adiabatic with respect to the $\ket{1,1}\leftrightarrow\ket{0,2}$ avoided crossing, produce phase gates
\begin{equation*}
U=\begin{pmatrix}
          1 & 0 & 0 & 0 \\
          0 & e^{i\phi_{01}} & 0 & 0 \\
          0 & 0 & e^{i\phi_{10}} & 0 \\
          0 & 0 & 0 & e^{i\phi_{11}}
        \end{pmatrix}
\end{equation*}
in the computational Hilbert space. Here, $\phi_{lr}=2\pi \int\delta\!f_{lr}(t)\,\mathrm{d}t$ is the dynamical phase acquired by $\ket{l,r}$, and
$\delta\!f_{lr}$ is the deviation of $f_{lr}$ from its value at point~I\@. A $\VR$ pulse into point~II
such that $\int\zeta(t)\, \mathrm{d}t=(2 n +1)\pi$ with integer $n$ implements a c-Phase, because
$\phi_{11}=\phi_{01}+\phi_{10}-\int\zeta(t)\,\mathrm{d}t$.
This method of realizing a c-Phase by adiabatically using the avoided crossing between computational and non-computational states
is generally applicable to any qubit implementation with finite anharmonicity, such as transmons~\cite{Schreier08} or phase qubits~\cite{Lucero08}.
A similar approach involving higher excitation levels but with non-adiabatic pulses was previously proposed~\cite{Strauch03}.
The negative anharmonicity permits the phase gate at point~II to occur before the onset of transverse coupling
at point~III.

Control of $\zeta$ by two orders of magnitude provides an excellent on-off ratio for the c-Phase gate. As shown in Fig.~\ref{fig:fig2}c, measurements of $\zeta$ obtained from  spectroscopy and from time-domain experiments show very good agreement. The time-domain method measures the  difference in the precession frequency of $\QL$  in two Ramsey-style experiments where a $\VR$-pulse of varying duration ($0$--$100\,\ns$) is  inserted between $\pi/2$ rotations of $\QL$,  with $\QR$ either in the ground state $\vert0\rangle$ or excited into state $\vert1\rangle$.
Using the time-domain approach, we measure a residual  $\zeta/2\pi\approx1.2\,\MHz$ at point~I (star).  The theoretical $\zeta$ obtained by numerical diagonalization shows reasonable agreement with the data, except for a scale factor that is likely due to higher modes of the cavity, not included in the calculation.

\begin{figure}[htpb!]
\psfrag{A}[c][c][1.0]{$\Rypt$}
\psfrag{B}[c][c][1.0]{$\cPhij$}
\psfrag{C}[c][c][1.0]{$\Rtomo$}
\psfrag{D}[c][c][1.0]{$\beta_1 \sigma_{z}^{\mathrm{L}} + \beta_2 \sigma_{z}^{\mathrm{R}}$}
\psfrag{E}[c][c][1.0]{$+\beta_{12} \sigma_{z}^{\mathrm{L}}\otimes\sigma_{z}^{\mathrm{R}}$}
\psfrag{F}[c][c][1.0]{$|\Psi^{+}\rangle = \frac{1}{\sqrt{2}}\left(|0,0\rangle+|1,1\rangle\right)$}
\psfrag{G}[c][c][1.0]{$|\Psi^{-}\rangle = \frac{1}{\sqrt{2}}\left(|0,0\rangle-|1,1\rangle\right)$}
\psfrag{H}[c][c][1.0]{$|\Phi^{+}\rangle = \frac{1}{\sqrt{2}}\left(|0,1\rangle+|1,0\rangle\right)$}
\psfrag{J}[c][c][1.0]{$|\Phi^{-}\rangle = \frac{1}{\sqrt{2}}\left(|0,1\rangle-|1,0\rangle\right)$}
\centering
\includegraphics[width=3.25in]{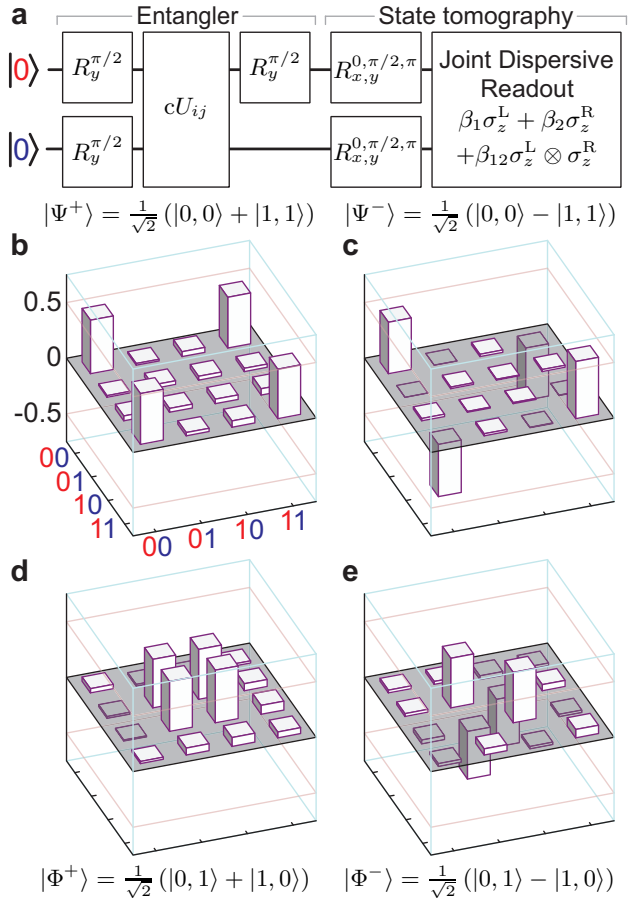}
\caption{{\bfseries Entanglement on demand}.
{\bfseries a},~Gate sequence generating two-qubit entanglement and detection via quantum state
tomography. Starting from $\ket{0,0}$, simultaneous $\pi/2$ rotations on both qubits create an equal superposition of the four computational states. A c-Phase $\cPhij$ then phase shifts $\ket{i,j}$ in the superposition and  produces entanglement. A final $\pi/2$ rotation on $\QL$ evolves the entangled state into one of the four Bell states depending on the $\cPhij$ applied.
{\bfseries b--e},~ Real part of maximum-likelihood density matrix $\rhoML$ of the entangler output for
$\cPhtw$, $\cPhz$, $\cPhth$, and $\cPho$, respectively (imaginary elements of $\rhoML$ are less than 0.03, 0.02, 0.07, 0.08).
Extracted metrics for the four entangler outputs include purity $P=0.87\pm0.02,0.92\pm0.02 ,0.88\pm0.02 ,0.79\pm0.03$, fidelity to the ideal Bell state $F=0.91\pm0.01, 0.94\pm0.01, 0.90\pm0.01, 0.87\pm0.02$ and concurrence $C=0.88\pm0.02, 0.94\pm0.01, 0.86\pm0.02, 0.81\pm0.04$. The uncertainties correspond to the standard deviation in 16 repetitions of generation-tomography for each entangler.
\label{fig:fig3}}
\end{figure}

The controlled phase interaction allows universal two-qubit gates. As an example, we produce high-fidelity entangled states on demand (Fig.~\ref{fig:fig3}). The pulse sequence in Fig.~\ref{fig:fig3}a generates any of the four Bell states,
\begin{equation*}
|\Psi^{\pm}\rangle = \frac{1}{\sqrt{2}}\left(|0,0\rangle \pm|1,1\rangle\right)\quad |\Phi^{\pm}\rangle = \frac{1}{\sqrt{2}}\left(|0,1\rangle \pm|1,0\rangle\right),
\end{equation*}
depending on the choice of c-Phase gate $\cPhij$ applied $(\cPhij\ket{l,r} = (-1)^{\delta_{il}\delta_{jr}} \ket{l,r})$. These gates
are realized through fine control of the dynamical phases $\phi_{01}$ and $\phi_{10}$ in a  $30\,\ns$ $\VR$-pulse close to point~II and back.
We tune $\phi_{01}$ over $2\pi$ by making small adjustments to the rising and falling edges of the $\VR$-pulse, and $\phi_{10}$ with a simultaneous weak $\VL$-pulse.

To detect the entanglement, we first reconstruct the two-qubit density matrix $\rho$ by quantum state tomography using joint dispersive readout~\cite{Blais04,Majer07,Filipp08}. A pulsed measurement of the cavity homodyne voltage  $\VH$ measures the operator
\begin{equation*}
	M=\beta_1\sigma_{z}^{\mathrm{L}} + \beta_2\sigma_{z}^{\mathrm{R}} + \beta_{12}\sigma_{z}^{\mathrm{L}}\otimes\sigma_{z}^{\mathrm{R}},
\end{equation*}
where the $\sigma$ are two-qubit Pauli operators~\cite{Nielsen00}. Operation in the
strong-dispersive regime~\cite{Schuster07,Filipp08} makes $|\beta_{12}|\sim |\beta_1|, |\beta_2|$, enhancing sensitivity to two-qubit correlations. A complete set of 15 linearly independent measurement operators is built using single-qubit rotations prior to measuring $M$. An ensemble average of each operator is obtained by executing the sequence in Fig.~\ref{fig:fig3}a  450,000 times. The 15 measured values are then input to a maximum likelihood estimator\cite{James01a} of $\rho$ (see Supplementary Information).

The inferred density matrices $\rhoML$  reveal highly-entangled states in all four cases (Fig.~\ref{fig:fig3}b--e).
We quantify performance using the metrics of  purity, $P(\rho)=\tr(\rho^2)$, fidelity to the target state $\ket{\psi}$,
$F(\rho,\psi)= \langle \psi | \rho | \psi \rangle$,
and concurrence~\cite{Wootters98}, $C$, computable from $\rhoML$.
Note that there are several common definitions of fidelity in the literature, and our definition is the square of the fidelity used in Refs.~\onlinecite{Steffen06} and \onlinecite{Filipp08}. Values for $P$, $F$ and $C$ for the
four cases are given in the caption to Fig.~\ref{fig:fig3}.  These values significantly extend the current state of the art for solid-state entanglement~\cite{Steffen06}, and provide evidence that we have a high-fidelity universal set of two-qubit gates.

\begin{figure}[tp!]
\psfrag{A}[c][c][1.0]{$\Rypt$}
\psfrag{B}[c][c][1.0]{$\Rypt$}
\psfrag{C}[c][c][1.0]{$O$}
\psfrag{D}[c][c][1.0]{$\Rypt$}
\psfrag{E}[c][c][1.0]{$\Rypt$}
\psfrag{F}[c][c][1.0]{$\cPhz$}
\psfrag{G}[c][c][1.0]{$\Rypt$}
\psfrag{H}[c][c][1.0]{$\Rypt$}
\psfrag{J}[c][c][1.0]{$\Rtomo$}
\psfrag{K}[c][c][1.0]{$\Rtomo$}
\centering
\includegraphics[width=3.25in]{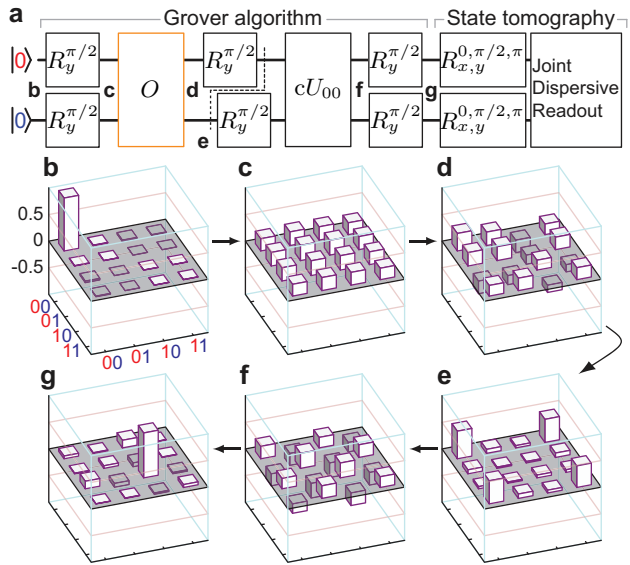}
\caption{{\bfseries Implementation of Grover's search algorithm}.
{\bfseries a}, Concatenation of single-qubit and c-Phase gates implementing one iteration of Grover searching. Without loss of generality,
we have replaced the Walsh--Hadamard transformations  $W=\Rxp\Rypt$ in the usual description of the algorithm~\cite{Nielsen00, Kaye07}
with $\Rypt$ rotations in order to eliminate 6 single-qubit rotations and complete the sequence in $104\,\ns$. (Supplementary Fig.~S3 shows
the microwave and flux pulses implementing the sequence.)
The orange box is the oracle $O=\cPhij$ that encodes the solution $x_0=2i+j$ to the search problem in a quantum phase. Note that the first half of the algorithm is identical to the entangling sequence in Fig.~\ref{fig:fig3}, while the second half is essentially its mirror image.
{\bfseries b--g}, Real part of $\rhoML$ obtained by state tomography after each  step of the algorithm with oracle $O=\cPhtw$.
Starting from $\ket{0,0}$ (b), the qubits are simultaneously rotated into a maximal superposition state (c). The oracle then marks
the solution, $\ket{1,0}$, by inverting its phase (d). The $\Rypt$ rotation on $\QL$ turns the state into the Bell state $\lvert\Psi^
+\rangle$, demonstrating that the state is highly entangled at this stage. The $\Rypt$ rotation on $\QR$ produces a state identical to (d) (data not shown). The application of $\cPhz$ undoes the entanglement, producing a  maximal superposition state (f). The final rotations yield an output state (g) with fidelity $F=85\%$ to the correct answer, $\ket{1,0}$.
\label{fig:fig4}}
\end{figure}

One- and two-qubit gates can be concatenated to realize simple algorithms, such as Grover's quantum search~\cite{Grover97} shown in Fig.~\ref{fig:fig4}. Given a function $f(x)$ on the set $x\in\{0,1,2,3\}$ such that $f(x)=1$ except at some $x_0$, where $f(x_0)=-1$, this well-known algorithm can determine $x_0=2i+j$ with a single call of an oracle $O=\cPhij$, which encodes $f(x)$ in a quantum phase.

We can examine the functioning of the algorithm by interrupting it after each step and performing state tomography. Figure~\ref{fig:fig4}b--g clearly shows all the features of a quantum processor, namely the use of maximally superposed states to exploit quantum parallelism (Fig.~\ref{fig:fig4}c), the encoding of information in the entanglement between qubits (Fig.~\ref{fig:fig4}d,~e), and the interference producing an answer represented in a final computational basis state. The fidelity of the final state (Fig.~\ref{fig:fig4}g) to the expected output ($\ket{1,0}$ for the case  $O=\cPhtw$ shown) is $85\%$. Similar performance is obtained for the other three oracles (Table~\ref{tab:summary}).\looseness=-1

We have also programmed and executed the Deutsch--Jozsa algorithm~\cite{Deutsch92,Cleve98}. The two-qubit version of this algorithm determines whether an unknown function $f_i(x)$, mapping a one-bit input to a one-bit output, is constant ($f_0(x)=0$ or $f_1(x)=1$) or balanced ($f_2(x)=x$ or $f_3(x)=1-x$), doing so with a single call of the function. The algorithm applies the function once to a superposition of the two possible inputs and employs the concept of quantum phase kick-back~\cite{Kaye07} to encode the result in the final state of one qubit (here, $\QL$) while leaving the other untouched ($\QR$). The gate sequence realizing the algorithm and the output tomographs for the four cases are shown in
Supplementary Fig.~S1.

The performance of both algorithms is summarized in Table~\ref{tab:summary}. Although there are undoubtedly significant systematic errors remaining, the overall fidelity is nonetheless similar to that expected from the ratio $(\sim100\,\ns/1\,\us)$ of the total duration of gate sequences to the qubit coherence times.

In summary, we have demonstrated the experimental realization of two-qubit quantum algorithms using a superconducting circuit.
The incorporation of local flux control and joint-dispersive readout into cQED, together with a ten-fold increase in
qubit coherence over previous two-qubit devices, has enabled on-demand generation and detection of entanglement and the
implementation of the Grover and Deutsch--Jozsa algorithms. Superconducting circuits could eventually perform more complex quantum algorithms on many qubits, provided that coherence lifetimes and the resulting gate fidelities can be further improved.

\section{Methods}
\subsection{Device fabrication}
A $180\,\nm$ film of Nb was dc-magnetron sputtered on the epi-polished surface of an R-plane corundum ($\alpha$-$\Sapphire$) wafer (2" diameter, $430\,\um$ thickness). Coplanar waveguide structures (cavity and flux-bias lines) were  patterned by optical lithography and
fluorine-based reactive ion etching of Nb.  Transmon features (interdigitated capacitors and split junctions) were patterned on individual
$2\, \mm \times7\,\mm$ chips using electron-beam lithography,  double angle evaporation of Al ($20/90\,\nm$) with intermediate oxidation
($15\%\,\mathrm{O}_2$ in Ar at $15\,\mathrm{Torr}$ for $12\,\mathrm{min}$), and lift-off.

A completed device was cooled to $13\,\mK$ in a $\Hethree$-$\Hefour$ dilution refrigerator.
A diagram of the refrigerator wiring is shown in Supplementary Fig.~S2.
Careful microwave engineering of the sample holder and on-chip wirebonding
across ground planes were required to suppress spurious resonance modes on- and off-chip.
Simulations using $\mathrm{Sonnet}^\circledR$ software provided guidance with this iterative process.
The sample was enclosed in two layers of Cryoperm magnetic shielding, allowing high-fidelity operation of the processor during
unattended overnight runs. \smallskip

\begin{table*}[ht!]
\topcaption{\textbf{Summary of algorithmic performance.}\label{tab:summary}}
\includegraphics[width=5in]{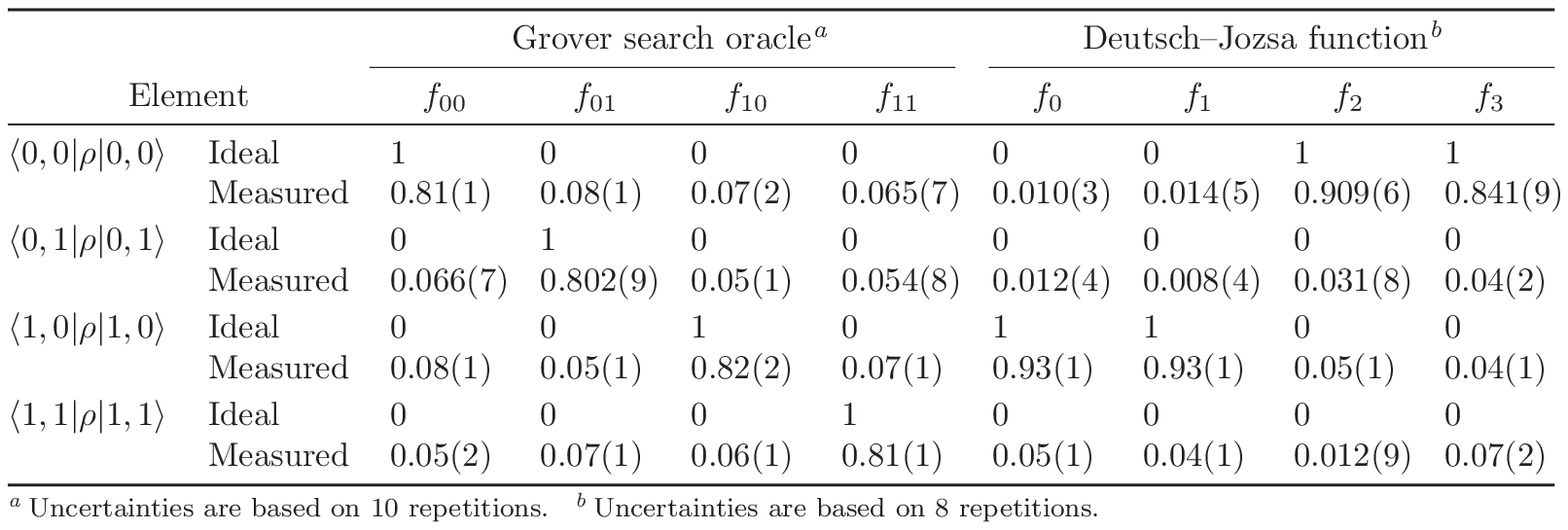}\\
\flushing
 Fidelity of the reconstructed output states of the Grover and Deutsch--Jozsa algorithms to their ideal outputs. These results suggest that, if combined with single-shot readout, the two algorithms executed with this processor would give the correct answer with probability far exceeding the $50\%$ success probability of the best classical algorithms limited to single calls of the oracle~\cite{Brickman05} or function.
\end{table*}

\subsection{cQED Theory}
The Tavis-Cummings~\cite{Tavis68} Hamiltonian generalized to multi-level transmon qubits~\cite{Koch07} is
\begin{eqnarray}
\label{eq:Ham}
H &=&\wcav a^{\dag}a + \\
  & &\sum_{q\in\{\!\mathrm{L},\mathrm{R}\!\}} \Bigl( \sum_{j=0}^{N} \omega_{0j}^q \lvert j\rangle_q\langle j\rvert_q + (a+a^{\dag}) \!\!\sum_{j,k=0}^{N} g_{jk}^{q}\lvert j\rangle_q\langle k\rvert_q \Bigr). \nonumber
\end{eqnarray}
Here, $\wcav$ is the bare cavity frequency, $\omega^q_{0j}=\omega_{0j}(\ECq, \EJq)$  is the transition frequency for qubit $q$ from ground to excited state $j$, and $g_{jk}^q=g_q n_{jk}(\ECq, \EJq)$, with  $g_q$ a bare qubit-cavity coupling and $n_{jk}$ a level-dependent coupling matrix element. The dependence of these parameters on qubit charging energy $\ECq$ and Josephson energy $\EJq$ is indicated. The flux control enters through $\EJq=\EJqmax\lvert\cos(\pi\Phi_q/\Phi_0)\rvert$, with  $\Phi_q$ the flux through the qubit loop, and a linear flux-voltage relation $\Phi_q=\alpha_{q\mathrm{L}} \VL + \alpha_{q\mathrm{R}}\VR +\Phi_{q,0}$, accounting for crosstalk ($\sim30\%$) and offsets. The above parameters are tightly constrained by the combination of spectroscopy and transmission data shown (Figs.~\ref{fig:fig1}b, \ref{fig:fig2}a and \ref{fig:fig2}b) and transmission data (not shown) for the $\QL$-cavity vacuum Rabi splitting. By simultaneously fitting the spectra given by numerical diagonalization of the Hamiltonian (truncated to  $N=5$ qubit levels and 5 cavity photons) to these data, we obtain $E_{\mathrm{JL(R)}}^{\mathrm{max}}/h= 28.48\,(42.34)\,\GHz$, $E_{\mathrm{CL(R)}}/h= 317\,(297)\,\MHz$,  $g_{\mathrm{L(R)}}/2\pi = 199\,(183)\,\MHz$,
and $\wcavontpi = 6.902\,\GHz$. The cavity linewidth is $\kappa/2\pi = 1\,\MHz$.\looseness=-1

\smallskip\noindent
\textbf{Acknowledgements} We thank V.~Manucharyan and E.~Boaknin for experimental contributions,
and M.~H.~Devoret, I.~L.~Chuang and A.~Nunnenkamp for discussions.
This work was supported by LPS/NSA under ARO Contract No.\ W911NF-05-1-0365,
and by the NSF under Grants No.\ DMR-0653377 and No.\ DMR-0603369.
We acknowledge additional support from CIFAR, MRI, MITACS, and NSERC (JMG),
NSERC, CIFAR, and the Alfred~P.~Sloan Foundation (AB), and
from CNR-Istituto di Cibernetica, Pozzuoli, Italy (LF).

\bibliography{References_Entanglement}

\end{document}

\title{Supplementary Material for\\ `Demonstration of Two-Qubit Algorithms with a Superconducting Quantum Processor'}
\author{L.\ DiCarlo}
\affiliation{Departments of Physics and Applied Physics, Yale University, New Haven, CT 06511, USA}
\author{J.\ M.\ Chow}
\affiliation{Departments of Physics and Applied Physics, Yale University, New Haven, CT 06511, USA}
\author{J.\ M.\ Gambetta}
\affiliation{Department of Physics and Astronomy and Institute for Quantum Computing, University of Waterloo, Waterloo, Ontario N2L 3G1, Canada}
\author{Lev\ S.\ Bishop}
\affiliation{Departments of Physics and Applied Physics, Yale University, New Haven, CT 06511, USA}
\author{B.\ R.\ Johnson}
\affiliation{Departments of Physics and Applied Physics, Yale University, New Haven, CT 06511, USA}
\author{D.\ I.\ Schuster}
\affiliation{Departments of Physics and Applied Physics, Yale University, New Haven, CT 06511, USA}
\author{J.\ Majer}
\affiliation{Atominstitut der \"{O}sterreichischen Universit\"{a}ten, TU-Wien, A-1020 Vienna, Austria}
\author{A.\ Blais}
\affiliation{D\'{e}partement de Physique, Universit\'{e} de Sherbrooke, Sherbrooke, Qu\'{e}bec J1K 2R1, Canada}
\author{L.\ Frunzio}
\affiliation{Departments of Physics and Applied Physics, Yale University, New Haven, CT 06511, USA}
\author{S.\ M.\ Girvin }
\affiliation{Departments of Physics and Applied Physics, Yale University, New Haven, CT 06511, USA}
\author{R.\ J.\ Schoelkopf}
\affiliation{Departments of Physics and Applied Physics, Yale University, New Haven, CT 06511, USA}
\date{May 1, 2009}
\maketitle

\section{Supplementary information}

\subsection{Perturbation Theory}

To gain additional insight on the large on-off ratio of the frequency shift $\zeta$, observed both in experiment and
numerical diagonalization of the Hamiltonian (Fig.~2),
we perform a perturbative analysis in the rotating-wave approximation.
 Truncating Eq.~(1) at three transmon excitations and assuming
$n_{12}\simeq \sqrt{2}$ (valid for $\EJ/\EC\gg1$), we obtain the fourth-order result
\begin{equation*}
\label{eq:pert}
\zeta=
-2 \gL^2 \gR^2 \left(  \frac{1}{\delta_1^{\phantom2} \DeltaL^2}+\frac{1}{\delta_2^{\phantom2}\DeltaR^2}+\frac{1}{\DeltaL^{\phantom2} \DeltaR^2}+\frac{1}{\DeltaR^{\phantom2}\DeltaL^2} \right).
\end{equation*}
Here, $\Delta_q=\omega_{01}^q-\wcav$, $\delta_1=\omega_{01}^{\mathrm{R}}-\omega_{12}^{\mathrm{L}}$, and $\delta_2=\omega_{01}^{\mathrm{L}}-\omega_{12}^{\mathrm{R}}$. This expression diverges  as the $0\leftrightarrow1$ transition of one transmon lines up with the $1\leftrightarrow2$ transition of the other.
Assuming instead two-level qubits, the expression simplifies to
\begin{equation*}
\label{eq:pertbad}
\zeta=
-2 \gL^2 \gR^2 \left(\frac{1}{\DeltaL^{\phantom2} \DeltaR^2}+\frac{1}{\DeltaR^{\phantom2}\DeltaL^2} \right).
\end{equation*}
Both perturbative expressions are compared with numerical diagonalization of the Hamiltonian in Fig.~2c. The three-level expression
shows reasonable agreement away from the divergence, while the two-level expression is incorrect in both magnitude and sign.

\subsection{State Tomography}
\label{sub:state_tomography}

The goal of quantum state tomography is to estimate the density matrix $\rho$ describing a quantum mechanical state.
For any two-qubit quantum state we can choose a set of $16$ linearly independent operators $\{M_i\}$ such that $\rho$ can be decomposed as
\begin{equation*}
	\rho = \sum_{i=1}^{16} c_i M_i,
\end{equation*}
where the set $\{c_i\}$ are the $16$ parameters to be estimated.
If the operators are observables, then the $16$ expectation values $m_i= \mathrm{Tr}[M_i\rho]$ determine $c_j$ by
\begin{equation*}\label{eq:tomography}
	m_i= \sum_{j=1}^{16} \mathrm{Tr}[M_iM_j] c_j.
\end{equation*}

Previous work~\cite{Wallraff05} has shown that in cQED a homodyne measurement of the cavity  is a faithful measurement of $\sigma_z$. For a quantum bus with two qubits the  measurement operator~\cite{Filipp08} is
\begin{equation*}
	M = \frac{1}{\tavg}\int_0^{\tavg}\!\! Q(t)\,\mathrm{d}t = \beta_1\sigma_{z}^{\mathrm{L}} + \beta_2\sigma_{z}^{\mathrm{R}} + \beta_{12}\sigma_{z}^{\mathrm{L}}\otimes\sigma_{z}^{\mathrm{R}}.
\end{equation*}
Here, $Q$ is the measured quadrature amplitude, $\tavg$ is an averaging window, and the $\beta$  are calibrated coefficients.
For this experiment, $\tavg= 450\,\ns$ and $(\beta_1,\beta_2,\beta_{12})\approx (60, 50, 40)\,\uV$.

Since the measurement contains both one- and two-qubit operators, a complete set of linearly independent operators $M_i$  can be made by applying only single-qubit rotations prior to measurement.
The set of 15 pre-rotations used in this experiment is  all combinations of $I$, $\Rxp$, $\Rxpt$, $\Rypt$ on left and right qubits, except that
$\Rxp \otimes \Rxp$ is not used. Only 15 measurements are needed to determine $\rho$ because of the constraint of trace normalization, $\tr \rho=1$ (equivalently we choose $M_{16}=I$, which always gives $m_{16}=1$).

Experimental averages $m_i$ are obtained by recreating the quantum state (executing the gate array), pre-rotating and measuring 450,000 times.
While ideally $\rho$ could be obtained from the experimental $m_i$ by inversion of $\mathrm{Tr}[M_iM_j]$, this method pays no attention to the properties $\rho$ must have: hermiticity and positive semi-definiteness (trace normalization is included by the choice of decomposition). These physical constraints are automatically included by a parametrization
\begin{equation*}
	\rho = \frac{T^\dagger T}{\mathrm{Tr}[T^\dagger T]},
\end{equation*}
where $T$ is a lower triangular matrix \cite{James01a}. For two qubits,
\begin{equation*}
	T = \begin{pmatrix}
		t_1 &  0  &  0     & 0 \\
		t_5+it_6 & t_2 &  0     & 0 \\
		t_{11}+it_{12} & t_7+it_8 & t_3 & 0 \\
	t_{15}+it_{16} & t_{13}+it_{14} & t_9+it_{10}  & t_4
	\end{pmatrix}.
\end{equation*}
The $t_i$ are found by standard Maximum likelihood Estimation \cite{James01a} with a  likelihood function
\begin{equation*}
	\mathcal{L}=\sum_{i=1}^{16} \alpha_i(m_i-\mathrm{Tr}[M_i\rho])^2,
\end{equation*}
where the $\alpha_i$ are weighting factors. We weight all measurements equally since amplifier noise dominates the error in all the measurements.

\bibliography{References_Entanglement}

\onecolumngrid\par

\begin{figure}[b!]
\psfrag{A}[c][c][1.0]{$\Rypt$}
\psfrag{B}[c][c][1.0]{$\Rymt$}
\psfrag{C}[l][c][0.9]{$l$}
\psfrag{D}[r][c][0.9]{$l$}
\psfrag{E}[l][b][0.9]{$r$}
\psfrag{F}[r][b][0.9]{$r\oplus f_i(l)$}
\psfrag{G}[c][c][1.0]{$U_i$}
\psfrag{H}[c][c][1.0]{$\Rypt$}
\psfrag{I}[c][c][1.0]{$\Rypt$}
\psfrag{J}[c][c][1.0]{$\Rtomo$}
\psfrag{K}[c][c][1.0]{$\Rtomo$}
\psfrag{L}[cB][cB][1.0]{$f_0(x)=0$}
\psfrag{M}[cB][cB][1.0]{$f_1(x)=1$}
\psfrag{N}[cB][cB][1.0]{$f_2(x)=x$}
\psfrag{O}[cB][cB][1.0]{$f_3(x)=1-x$}
\centering
\includegraphics[width=3.25in]{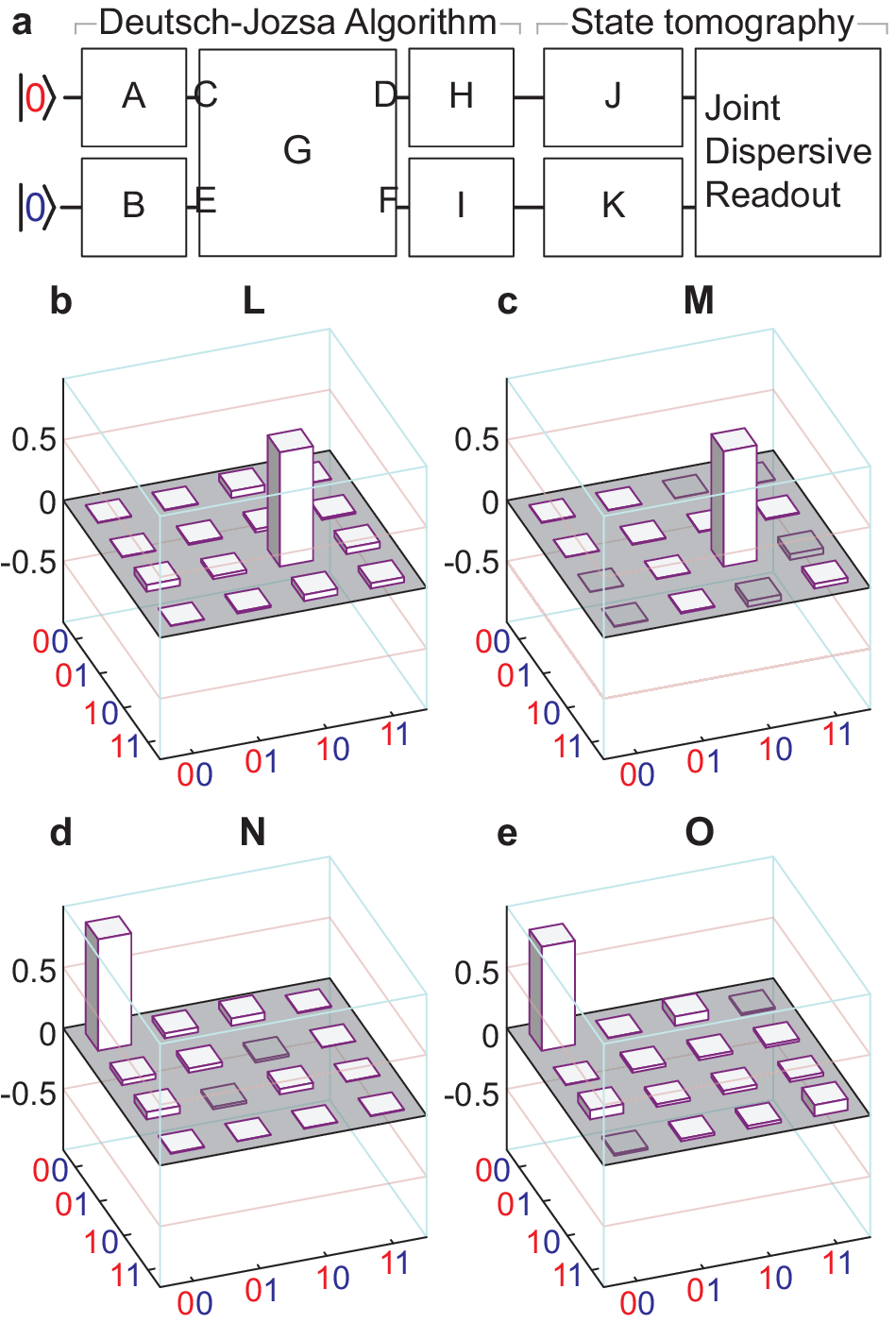}
\caption{{\bfseries Implementation of Deutsch--Jozsa algorithm}.
{\bfseries a}, Gate sequence solving the Deutsch--Jozsa problem.
The two-qubit gates $U_i$ performing the transformation
$\ket{l,r}\rightarrow \ket{l, r \oplus f_i(l)}$ ($\oplus$ denotes addition modulo 2) for $f_0(x)=0$, $f_1(x)=1$, $f_2(x)=x$, and $f_3(x)=1-x$ are  $U_0= \Id\otimes\Id$, $U_1=\Id\otimes\Rxp$, $U_2=(\Id\otimes\Rypt\Rxp)\cPhz(\Id\otimes \Rypt)$, and $U_3=(\Id\otimes \Rymt\Rxp)\cPhth(\Id\otimes \Rymt)$, respectively.
{\bfseries b--e}, Real part of the inferred density matrix $\rhoML$ of the algorithm output in the four cases
(imaginary elements of $\rhoML$ are less than 0.05, 0.03, 0.05, 0.06, respectively). For the constant (balanced) functions $f_0$ and $f_1$ ($f_2$ and $f_3$), $\rhoML$ reveals high fidelity to $\ket{1,0}$ ($\ket{0,0}$), as expected. For the tomographs shown, the fidelity to the ideal output state is $F=0.94$, 0.95, 0.92, and 0.85, respectively.  Statistics for 8 runs of each of the four cases are given in Table~I.
\label{fig:fig5}}
\end{figure}

\begin{figure*}[hp!]
\centering
\includegraphics[width=5.5in]{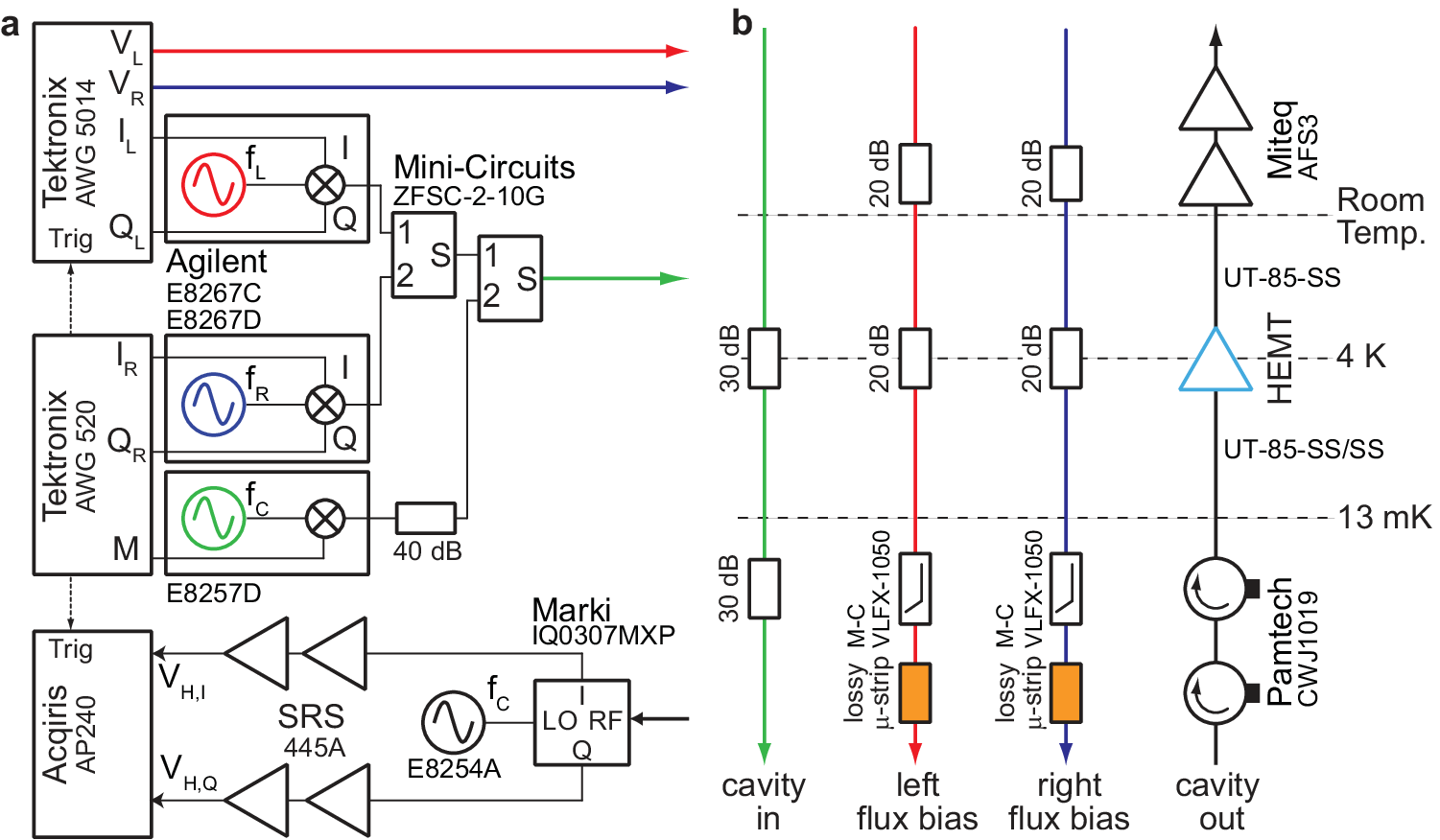}
\caption{{\bfseries Experimental setup and wiring}.
{\bfseries a}, Block diagram of room-temperature electronics. Arbitrary waveform generators, with $1\,\ns$ sampling rate and 10-bit resolution, produce voltages $\VL$ and $\VR$ directly applied to the flux-bias lines, the I-Q
modulation envelopes for the microwave tones driving single-qubit  $x$- and $y$-rotations, and the pulse that modulates the cavity measurement. On the output side, an I-Q mixer and a two-channel averager ($2\,\ns$, 8-bit sampling) complete the readout chain performing homodyne detection of the cavity quadratures. The arbitrary waveform generators, microwave synthesizers and acquisition card are clocked with a Rubidium frequency standard (SRS~FS725, not shown). {\bfseries b}, Schematic of the microwave wiring inside the dilution refrigerator, showing heavily-attenuated input lines and an output chain with $\sim100\,\dB$ gain in the $4$--$8\,\GHz$ range.
\label{fig:fig6}}
\end{figure*}

\begin{figure*}[hp!]
\psfrag{H}[c][c][.9]{$\Rypt$}
\psfrag{K}[c][c][.9]{$\Rxp$}
\psfrag{D}[c][c][.9]{$\Rxpt$}
\psfrag{E}[c][c][.9]{$\cPhtw$}
\psfrag{G}[c][c][.9]{$\cPhz$}
\psfrag{s}[c][c][.9]{$4\sigma$}
\centering
\includegraphics[width=3.25in]{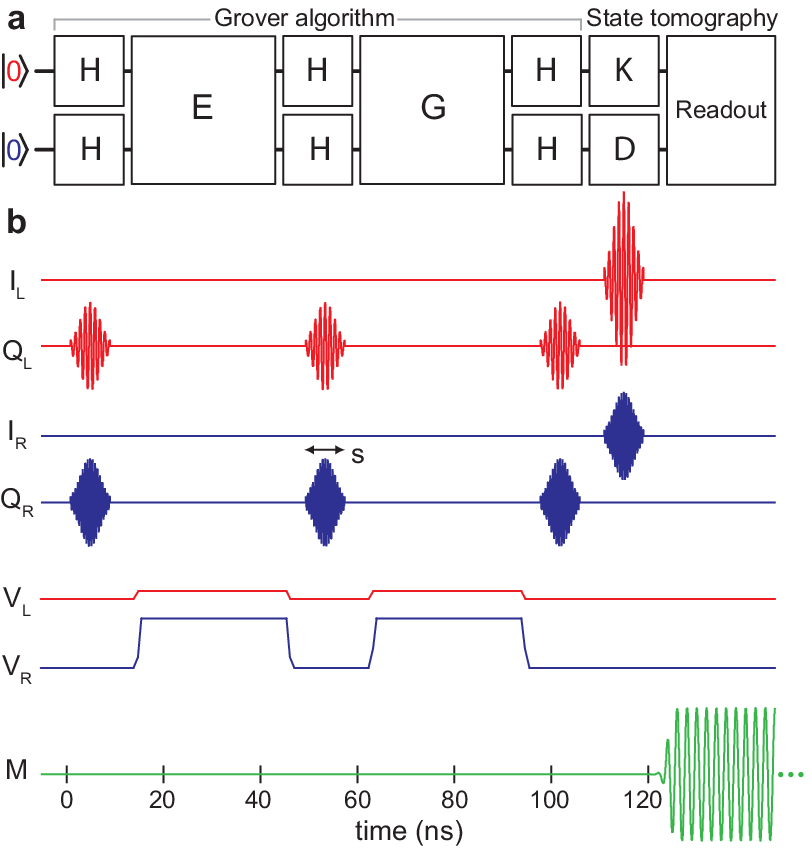}
\caption{{\bfseries Microwave and flux pulses realizing the gates}.
{\bfseries a}, An example sequence, executing the Grover search  algorithm with
oracle $O=\cPhtw$ and measuring $M_{13}=-\beta_1\sigma_{z}^{\mathrm{L}} + \beta_2\sigma_{y}^{\mathrm{R}} - \beta_{12}\sigma_{z}^{\mathrm{L}}\otimes\sigma_{y}^{\mathrm{R}}$.
{\bfseries b}, Illustration of the microwave and flux pulses realizing the operations directly above.
All microwave pulses implementing the $x$- and $y$-rotations have Gaussian envelopes, with standard
deviation $\sigma=2\,\ns$, truncated at $\pm 2 \sigma$. The rotation axis is set using I-Q (vector) modulation (see Fig.~\ref{fig:fig6}), and rotation angle is controlled by pulse amplitude. A buffer of $5\,\ns$ is inserted between all microwave and flux pulses to avoid any overlap.
\label{fig:fig7}}
\end{figure*}

\newpage